\documentclass[twocolumn,trackchanges]{aastex701}

\usepackage{bm}
\usepackage{comment}
\usepackage{amsmath}

\begin{document}

\title{Self-organization of local streamline structures and energy transfer rate in compressible plasma turbulence}

\author[0000-0002-7102-5032]{Simone Benella}
\affiliation{INAF - Institute for Space Astrophysics and Planetology, Via del Fosso del Cavaliere, 100, 00133, Rome, Italy}
\email[show]{simone.benella@inaf.it}  

\author[0000-0001-7288-1087]{Virgilio Quattrociocchi} 
\affiliation{INAF - Institute for Space Astrophysics and Planetology, Via del Fosso del Cavaliere, 100, 00133, Rome, Italy}
\email{virgilio.quattrociocchi@inaf.it} 

\author[0000-0002-7969-7415]{Emanuele Papini}
\affiliation{INAF - Institute for Space Astrophysics and Planetology, Via del Fosso del Cavaliere, 100, 00133, Rome, Italy}
\email{emanuele.papini@inaf.it} 

\author[0000-0003-4380-4837]{Andrea Verdini}
\affiliation{Dipartimento di Fisica e Astronomia, Università di Firenze, Via G. Sansone 1, Sesto Fiorentino (Firenze), 50019, Italy}
\email{andrea.verdini@unifi.it} 

\author[0000-0002-1322-8712]{Simone Landi}
\affiliation{Dipartimento di Fisica e Astronomia, Università di Firenze, Via G. Sansone 1, Sesto Fiorentino (Firenze), 50019, Italy}
\email{simone.landi@unifi.it} 

\author[0000-0002-5002-6060]{Maria Federica Marcucci}
\affiliation{INAF - Institute for Space Astrophysics and Planetology, Via del Fosso del Cavaliere, 100, 00133, Rome, Italy}
\email{maria.marcucci@inaf.it} 

\author[0000-0002-3403-647X]{Giuseppe Consolini}
\affiliation{INAF - Institute for Space Astrophysics and Planetology, Via del Fosso del Cavaliere, 100, 00133, Rome, Italy}
\email{giuseppe.consolini@inaf.it} 

\begin{abstract}

We examine how local streamline topology and energy cascade rate self-organize in plasma turbulence for both compressible and incompressible regimes. Using a fully-compressible Hall-magnetohydrodynamic simulation, we quantify the subgrid-scale energy transfer and analyze its relationship to streamline structures by means of grandient tensor geometric invariants of the velocity field. Our results highlight how streamline topology is crucial for diagnosing turbulence: for nearly-incompressible fluctuations the energy is primarily transferred to smaller scales through strain-dominated and stable-vortical structures, while is back-transferred towards larger scales through unstable-vortical structures. Compressible fluctuations, on the contrary, do not show a clear topological selection of the energy transfer since the overall direction of the local cascade rate is found to be determined by the sign of $-\nabla\cdot\bm{u}$ (plasma volumetric compression or expansion). %The identified connection between streamline topology and local energy transfer offers a more complete picture of turbulent cascade in space plasmas.

\end{abstract}


\section{Introduction} 
Turbulence in space plasmas remains a major open problem, despite significant advances from theory, improved spacecraft in situ measurements, and numerical simulations \citep{Bruno2016}. In a turbulent plasma, the large-scale energy budget is transferred across the magnetohydrodynamic (MHD) inertial range toward ion scales through a cascading process that generates a hierarchy of structures at all scales. Observational and numerical studies have shown that while the average direction of the energy transfer is predominantly forward, inverse energy transfer can also occur at all scales. In the case of incompressible plasmas, for example, the emergence of inverse cascade has been linked to the imbalance between counter propagating Alfvén waves, suggesting a dependence of the local energy transfer to the large-scale organization of the turbulent fluctuations \citep{Smith2009,Stawarz2010,Coburn2014}. In addition to the local energy transfer associated with nonlinear interactions, at ion scales kinetic instabilities and other nonlocal processes can significantly contribute to the energy transfer. For instance, magnetic reconnection occurring at sub-ion scales has been shown to induce an inverse energy transfer toward MHD scales \citep{Franci2017,Manzini2023,Foldes2024}. %At the end of the inertial range, plasma turbulence enters a transition region leading to kinetic scales, where both forward and inverse energy transfer between small scales and the ion inertial length $d_i$ can occur.
An heuristic proxy of the energy transfer rate, which is local in time (space), hence called \textit{local energy transfer} (LET), has been introduced by \cite{Sorriso2018a} in the context of numerical simulations. This quantity provides useful insights about the irregular and burst-like activity associated with the transfer rate from and towards small scales, and possibly constituting an indirect proxy of turbulent structures. The operative definition of the LET is based on the Yaglom's law \citep{Politano1998,Marino2023}, which is averaged on a fixed short time window, allowing to assess the locality of the energy transfer in time. The deep link between LET and kinetic plasma properties has been emphasized by \citep{Sorriso2019}, where the LET has been estimated on Magnetospheric Multiscale \citep[MMS;][]{Burch2016} measurements.

%The access to the three-dimensional fields in the space domain that we have on numerical simulations, allows us to calculate in a more robust way the energy transfer by resorting to the space-filter technique, firstly introduced in the context of hydrodynamic turbulence for large-eddy simulations \citep{Germano1992,Meneveau2000} and then extended to MHD \citep{Aluie2010,Aluie2017}. The space-filter indeed is as a coarse-graining (CG) of the fields which smooths the smallest-scale dissipative patterns. The generality and robustness of the space-filter approach relies on the filtering procedure, that can be directly applied on the dynamical equations. At finite scale, this filtering produces extra terms in the CG equations which couple different scales that represent the subgrid-scale (SGS) energy exchange. From the mathematical point of view the SGS energy transfer terms have been shown to be equivalent to the LET \citep{Aluie2017,Manzini2022}. The great advantage is that with the SGS energy transfer we do not need to invoke the validity of the Yaglom's law, which holds in the inertial range. Thus, it represents a generalization of the LET and can be calculated at any scale in the system, even outside the MHD range of turbulence.

As energy flows across scales, turbulence organizes in a hierarchy of coherent, multiscale structures. An effective way of investigating the local properties of such behavior is based on the statistics of the gradient-tensor geometrical invariants (GTGIs). The first investigation of the properties and the lagrangian dynamics of GTGIs has been provided by \cite{Vieillefosse1982}, who introduced the evolution equation of GTGIs for a restricted Eurler system \citep[see also][]{Vieillefosse1984}. This simple theoretical framework has been later extended to more sophisticated models, such as shell models by \cite{Biferale2007}, \citep[see][for a review]{Meneveau2011}. The typical hallmark of energy dissipation in the statistics of the GTGIs consists of an asymmetric feature called the \textit{Vieillefosse tail}. In addition to the lagrangian time evolution of GTGIs, further investigations also elucidated the scale dependent behavior of GTGI statistics. \cite{Chertkov1999} showed that by coarse-graining the velocity field, the Vieillefosse tail tends to disappear when moving towards the inertial range. In the framework of MHD simulations, \citet{Dallas2013} investigated the statistics of GTGIs of velocity and magnetic fields by means of MHD direct numerical simulations providing an extension of earlier hydrodynamic theories.

First observational evidence of the presence of the Vieillefosse tail in the GTGI statistics in space plasmas was reported in \cite{Consolini2015} leveraging the multipoint measurements provided by the Cluster spacecraft constellation \citep{Escoubet1997}. In that work, authors showed that the typical asymmetry in the velocity-field gradients routinely observed in hydrodynamics is also present in the solar wind. Conversely, the absence of this feature in the magnetic field GTGIs, reported by \cite{Dallas2013} in MHD simulations, has been confirmed by numerous studies based on data gathered by Cluster and Magnetospheric Multiscale (MMS) missions \citep{Quattrociocchi2019,Hnat2021,Zhang2023,Quattrociocchi2025}. All these works provide important clues on the properties of turbulent systems by quantifying the statistics of the formation of local structures, but their relation to the energy transfer has not been fully assessed yet. Evidence of the existing correlation between the energy transfer rate and the GTGIs, which elucidates how the local topology of streamline structures connects to the local turbulent cascade, has been introduced in the context of hydrodynamics \citep{Bos2002,Vela2021,Yao2024}.
In the field of space plasmas, the crucial interplay between field line topology and LET has been assessed by \cite{Hnat2025} for the magnetic field. They show that positive LET (i.e., direct energy cascade) exhibits a significant correlation with the occurrence of strain-dominated magnetic field structures, suggesting the crucial role of fundamental processes, such as magnetic reconnection, in driving the energy cascade. %However, when moving towards kinetic scales, \cite{Bandyopadhyay2020} investigated the possible correlation between the pressure-strain interaction and the vortical structures, symmetric strain and currents as quadratic invariants in particle-in-cell simulations and MMS data. They highlight the importance of the velocity field topology in leading the energy dissipation, being the enhanced pressure-strain interaction co-localized with both vortical and strain-dominated structures of the ion and electron velocity GTGIs.
\begin{figure*}
    \centering
    \includegraphics[width=\linewidth]{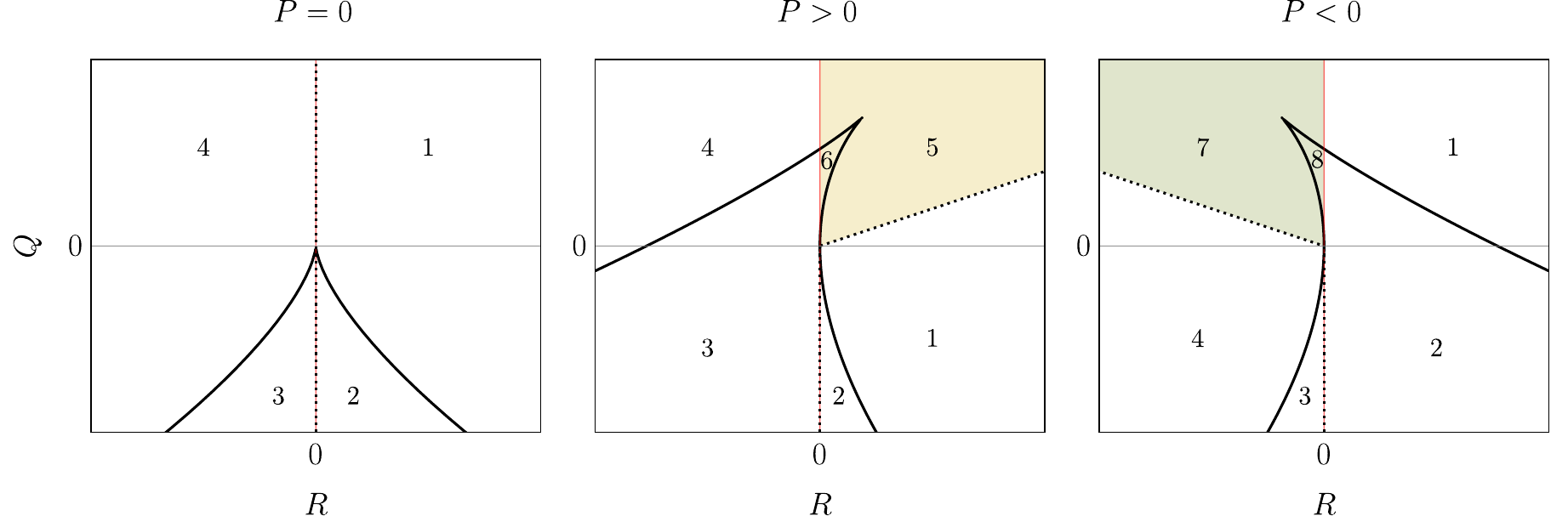}
    \caption{Sketch of the local streamline topologies in the $RQ$-plane for the three cases $P=0$ (left), $P>0$ (middle), and $P<0$ (right). Thick line is the discriminant surface $\Delta$. Red line marks the separation between stretching and compressing regions, and the dashed line indicate the separation between stable and unstable topologies. Shaded areas indicate topologies associated with local volumetric contraction (yellow) and expansion (green) emerging in compressible cases only.}
    \label{fig:pqr}
\end{figure*}
Another key aspect of all these studies presenting the statistics of GTGIs in space plasmas is that they rely on the assumption of incompressibility, which is encoded in the GTGI equations used in the data analysis. In this context, the aim of this letter is two-fold: to reveal the intimate relation between local streamline topology and energy transfer rate, and to investigate and compare this connection in both compressible and incompressible cases. All these steps are carried out by analyzing a three-dimensional Hall-MHD numerical simulation of plasma turbulence.

\section{The geometric invariants of the gradient tensors}

The method adopted in this work relies on the geometric invariants of the velocity field gradient tensor $\bm{A}=\nabla\bm{u}$. To define invariants quantities, we introduce the characteristic polynomial of the gradient tensor. Starting from the expression of this polynomial, it is possible to define
\begin{comment}
\begin{align}
\nonumber
    p(\lambda) &= \lambda^3 - \lambda^2 \mathrm{Tr}(\bm{A}) +\frac{\lambda}{2}((\mathrm{Tr}\bm{A})^2\\
    &-\mathrm{Tr}(\bm{A}^2)) - \mathrm{det}(\bm{A})
    \label{eq:poly}
\end{align}
Starting from Equation \eqref{eq:poly}, it is possible to define 
\end{comment}
characteristic quantities that are invariant under $SO(3)$-group transformations (i.e., rotations and reflections). These quantities are related to the trace and the determinant of the matrix. The three gradient tensor geometric invariants (GTGIs) thus are
\begin{align}
P &= -\mathrm{Tr}(\bm{A})=-(\lambda_1+\lambda_2+\lambda_3)\label{eq:p}\\
Q &= \frac{1}{2}(P^2 - \mathrm{Tr}(\bm{A}^2))=\lambda_1\lambda_2+\lambda_2\lambda_3+\lambda_3\lambda_1\\
R &= -\mathrm{det}(\bm{A}) = -\lambda_1\lambda_2\lambda_3,
\end{align}
where $\lambda_i, i=1\dots3$ are the eigenvalues of $\bm{A}$, which encode fundamental information on the local flow topology. Denoting by $\lambda_1$ the real eigenvalue, the remaining pair $\lambda_2,\lambda_3$ may be either real (pure strain) of complex conjugate (vortical motion). In the latter case, the sign of $\mathrm{Re}\lambda_{i=2,3}$ determines whether the local topology is stable ($\mathrm{Re}\lambda_{i=2,3}<0$) or unstable ($\mathrm{Re}\lambda_{i=2,3}>0$). When all the eigenvalues are real, the stability is instead expressed by the sign of $\lambda_2+\lambda_3$, i.e., the net contraction or expansion on the plane orthogonal to the principal direction.
The sign of the principal eigenvalue $\lambda_1$ indicates compressing ($\lambda_1<0$) or stretching ($\lambda_1>0$) topologies. For incompressible flows ($P=0$), the divergenceless costraint $\lambda_1=-(\lambda_2+\lambda_3)$ couples the dynamics along the principal direction to the transverse plane. Consequently, an unstable transverse topology ($\lambda_2+\lambda_3>0$) necessarily implies a compression with rate $-(\lambda_2+\lambda_3)$ along the principal direction, and vice versa. In other words, flow configurations including simultaneous contraction (``sinks'') or dilation (``sources'') along all the directions are forbidden. In the compressible flows ($P\neq0$), this constraint is relaxed and additional topologies, such as local volumetric expansion or contraction along all directions, become admissible \citep{Suman2010,Wang2012,Vaghefi2015,Zheng2022}. By considering the case of negative divergence ($P>0$), we can write the condition $\lambda_1<-(\lambda_2+\lambda_3)$, which implies that unstable transverse topologies are always associated with a compression stronger than $-(\lambda_2+\lambda_3)$ on the principal direction, thus precluding the formation of local sources. Conversely, for the positive divergence case ($P<0$), simultaneous contractions along all direction is locally forbidden, since the condition $\lambda_1>\lambda_2+\lambda_3$ holds in this case. In other words, stable transverse topologies are always coupled with a stretching rate larger than $\lambda_2+\lambda_3$ along the principal direction $\lambda_1$, effectively preventing the formation of local sinks.
%As for any generic gradient tensor, $\bm{A}$ can be decomposed as the sum of a symmetric and a skew-symmetric contribution as follows:
%\begin{equation}
%    \bm{A} =\bm{S} + \bm{\Omega}=S_{lm}-\frac{1}{2}\epsilon_{lmn}\omega_n,
%\end{equation}
%thus separating the strain tensor $\bm{S}=\frac{1}{2}(\nabla\bm{u} + \nabla\bm{u}^T)$ from the rotation rate tensor (i.e., the vorticity) $\bm{\Omega}=\frac{1}{2}(\nabla\bm{u} - \nabla\bm{u}^T)$. Using strain- and rotation-rate tensors, we can express the velocity gradient GTGIs as
%\begin{align}
%    P &=-\nabla\cdot\bm{u}\\
%    Q &=\frac{1}{4}[\omega^2-2\,\mathrm{tr}(\bm{S}^2)]\\
%    R &=-\frac{1}{3}[\mathrm{tr}(\bm{S}^3)+\frac{3}{4}\omega_l\omega_m S_{lm}].
%\end{align}
%where $\omega_i = \varepsilon_{ijk}\Omega_{jk}$. This allows us to readily observe that the first invariant is directly related to the divergence of $\bm{u}$ and $\bm{b}$, therefore it vanishes in the case of incompressible/solenoidal fields. The second invariant quantifies the balance between strain- and rotation-rate local topology of the field lines, whereas the third invariant is related to the stability of the topological structures. Indeed, it comprises the strain self-amplification term and the strain-rotation interplay, accounting for vortex stretching.

The surface that divides the real solutions from the complex ones is given by the discriminant line
\begin{equation}
    \Delta = 27R^2 + (4P^3 - 18PQ)R + (4Q^3 - P^2Q^2) = 0,
    \label{eq:d_p_neq_0}
\end{equation}
which in the incompressible case reduces to
\begin{equation}
    \Delta|_{P=0} = 4Q^3 + 27R^2=0.
    \label{eq:d_p_eq_0}
\end{equation}
This surface separates the $RQ$-plane in two parts (see Figure \ref{fig:pqr}, left panel), the $\Delta|_{P=0} > 0$ region where we find vortical compressing (1) and stretching (4) structures (one real and two complex-conjugate eigenvalues), and the $\Delta|_{P=0} < 0$ region, where the streamline are subject to pure compressing (2) or stretching (3) strain (three real eigenvalues). The compression/stretching separation in the $R=0$ line, that in the incompressible case coincides with the stable/unstable separatrix $PQ-R=0$. When $P>0$, in addition to the topologies from (1) to (4), the simultaneous modification of the $\Delta$ surface along with the stable/unstable separatrix that deviates from the line $R=0$, induces two new topologies: stable compressing vortical (5) and strain (6) structures, i.e., local sinks (Figure \ref{fig:pqr} middle panel). Analogous considerations explain the emergence of two opposite topologies in the case $P<0$ (Figure \ref{fig:pqr}, right panel) which are unstable stretching vortical (7) and strain (8) structures, thus acting as local sources \citep{Vaghefi2015}.
%\begin{figure}
%    \centering
%    \includegraphics[width=0.8\linewidth]{figures/rq_sketch.pdf}
%    \caption{A sketch of the $RQ$-plane along with the typical local topology of the field lines. The magenta line indicate the discriminant separatrix $\Delta=0$.}
%    \label{fig:rq_sketch}
%\end{figure}

\section{Coarse-graining and energy transfer rate}\label{sec:3}

To investigate the local energy cascade rate across a scale $\ell$, we introduce a coarse-graining (CG) based on the space-filter approach \citep{Germano1992,Eyink2005,Aluie2010,Aluie2017,Camporeale2018,Manzini2022,Foldes2024}, where the CG consists of a Gaussian kernel $G_\ell(\bm{x})=\ell^{-3}G(\bm{x}/\ell)$ with zero mean and unit variance, i.e.,
\begin{equation}
    \bar{f}(\bm{x}) = \int d\bm{r}G_\ell(\bm{r})f(\bm{x}+\bm{r}).
    \label{eq_gauss_filt}
\end{equation}
Since we are interested in analyzing a compressible simulation, for the velocity field we use the Favre average $\tilde{u}(\bm{x})=\overline{\rho u}(\bm{x})/\bar{\rho}(\bm{x})$, \citep{Favre1975,Aluie2011}.
The CG of the field $f$ at a scale $\ell$ results in two contributions: the large scale field $\bar{f}_\ell$, and the sub scale field $f^{'} = f-\bar{f}_\ell$, the latter being the subgrid-scale (SGS) unresolved part of the original field $f$. By applying the CG to the set of Hall-MHD equations in Alfvén units \citep{Aluie2011,Aluie2017,Manzini2022}, we obtain the following SGS energy transfer terms (see Appendix \ref{app:hmhd_cg}):
%\begin{eqnarray}
%    \partial_t \bm{u} &=& -(\bm{u}\cdot\nabla)\bm{u} + (\bm{b}\cdot\nabla)\bm{b} - \nabla P +\bm{d}_{\nu}+\bm{f}\\
%    \partial_t \bm{b} &=& \nabla \times (\bm{u}\times\bm{b}) -d_i \nabla \times (\bm{j}\times \bm{b}) + \bm{d}_{\eta}
%    \label{HMHD}
%\end{eqnarray}
%The CG operation introduced in Equation \eqref{CG} is a convolution, so it commutes with space and time derivatives, and to obtain the filtered equations at the scale $\ell$ we have to convolve Equation \eqref{HMHD} with the Gaussian kernel $G_\ell$. If we \textit{filter} the last equations at the scale $\ell$, we obtain the following expressions:
%\begin{align}
%    \partial_t \tilde{\bm{u}} &= -(\tilde{\bm{u}}\cdot\nabla)\tilde{\bm{u}} + (\tilde{\bm{b}}\cdot\nabla)\tilde{\bm{b}} - \nabla \tilde{P} + \tilde{\bm{d}}_{\nu} - \nabla \cdot \bm{\tau} + \tilde{\bm{f}}\nonumber\\
%    \partial_t \tilde{\bm{b}} &= \nabla \times[(\tilde{\bm{u}}-d_i \tilde{\bm{j}})\times\tilde{\bm{b}}] + \tilde{\bm{d}}_{\eta} + \nabla \times (\mathcal{E}_{M} + d_i \mathcal{E}_{H}),
%    \label{eq:hmhd_cg}
%\end{align}
\begin{align}
    \pi_{sgs}^\text{RM}=&-[\bar{\rho}\,\tilde{\tau}(u_j,u_k)-\bar{\tau}(b_j,b_k)]\,\partial_k\tilde{u}_j\\
    \pi_{sgs}^\text{BPY}=&-\frac{1}{\bar{\rho}}\,\partial_j \bar{p}\,\bar{\tau}(\rho,u_j)\\
    \pi_{sgs}^\text{MHD}=&-\mathcal{E}^\text{MHD}_j\,\bar{J}_j\\
    \pi_{sgs}^\text{Hall}=&-d_i\mathcal{E}^\text{Hall}_j\,\bar{J}_j,
\end{align}
where $\rho$ is the density, $\bar{\tau}(f,g)=\overline{f g} - \bar{f} \bar{g}$ is the SGS stress tensor at scale $\ell$, $b_k$ is the magnetic field, $p$ is the plasma pressure, $J_k$ is the current density, $d_i$ is the ion inertial length, $\mathcal{E}^\text{MHD}_k=\epsilon_{jkl}(\overline{u_kb}_l-\tilde{u}_k\bar{b}_l)$ and $\mathcal{E}^\text{Hall}_k=\epsilon_{jkl}(\overline{J_kb_l/\rho}-\bar{J}_k\bar{b}_l/\bar{\rho})$ are the MHD and Hall electromotive forces, respectively. The Reynolds-Maxwell (RM), MHD, and Hall energy transfer terms are the same of the incompressible case \citep[e.g., see][]{Manzini2022} with appropriate corrections due to density fluctuations (i.e., Favre filtering and $\bar{\rho}^{-1}$ factor in the Hall term). The baropycnal (BPY) work, instead, is a compressible-specific term due to the action of large-scale pressure gradients on turbulent mass fluxes \citep[e.g., see][]{Aluie2011}.
%where $p$ is the plasma pressure, $\bm{d}_\nu=\nu\nabla^2\bm{u}$, $\bm{f}$ is a generic forcing, $d_i$ indicates the ion inertial length, $\bm{j}$ is the current density, $\bm{d}_\eta=\eta\nabla^2\bm{b}$, and $\mathcal{E}_M$ and $\mathcal{E}_H$ are the MHD and Hall electric field terms, respectively.
%We have also introduced the quantity $\bm{\tau}$ that corresponds to the Reynolds-Maxwell stress tensor $\tau_{ij} = \tau (u_i;u_j) - \tau(b_i;b_j) = \tilde{u_iu_j} - \tilde{u}_i\tilde{u}_j - (\tilde{b_ib_j}-\tilde{b}_i\tilde{b}_j)$.

\section{Results}

\begin{figure}
    \centering
    \includegraphics[width=\linewidth]{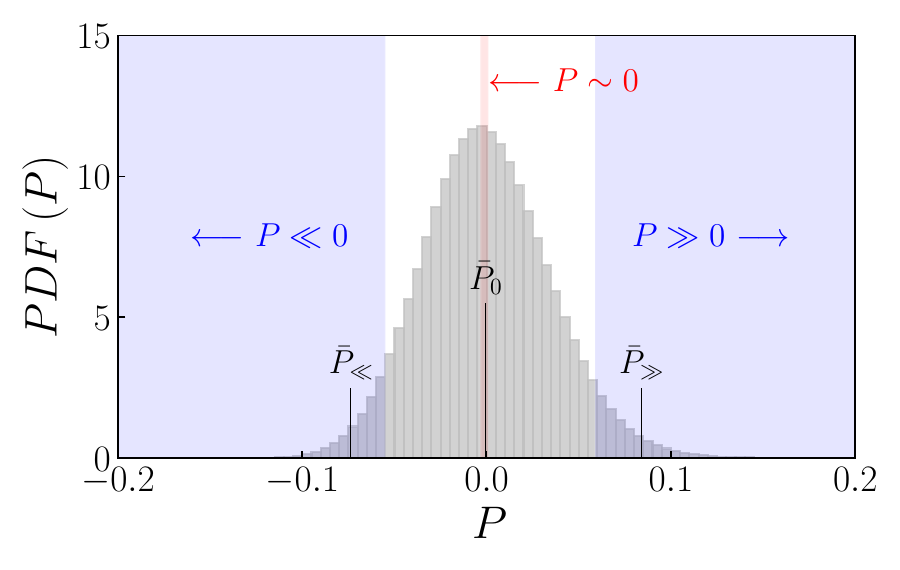}
    \caption{PDF of the first GTGI $P=-\nabla\cdot\bm{u}$. Shaded areas indicate $P\ll0$, $P\gg0$ (blue), and $P\sim0$ (red) regimes. Vertical lines indicate the median values $\bar{P}_\ll=-0.06$, $\bar{P}_0=-1.2\times10^{-3}$, and $\bar{P}_\gg=0.07$ inside each subset.}
    \label{fig:p_stat}
\end{figure}
%
%\begin{figure*}
%    \centering
%    \includegraphics[width=0.49\linewidth]{figures/pPI_1.pdf}
%    \includegraphics[width=0.49\linewidth]{figures/pP_PI_1.pdf}
%    \caption{PDF of the first GTGI $P=-\nabla\cdot\bm{u}$. Shaded areas indicate $P\ll0$, $P\gg0$ (blue), and $P\sim0$ (red) regimes. Vertical lines indicate the median values $\bar{P}_\ll=-0.06$, $\bar{P}_0=-7\times10^{-4}$, and $\bar{P}_\gg=0.07$ inside each subset.}
%    \label{fig:p_stat}
%\end{figure*}
%
\begin{figure*}
    \centering
    \includegraphics[width=0.49\linewidth]{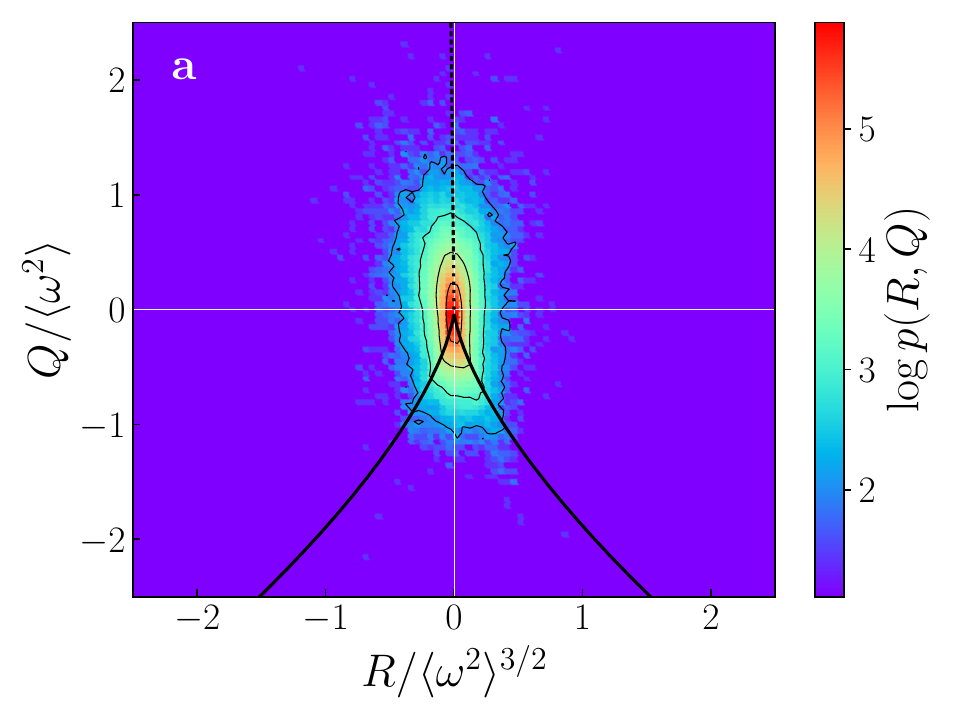}
    \includegraphics[width=0.49\linewidth]{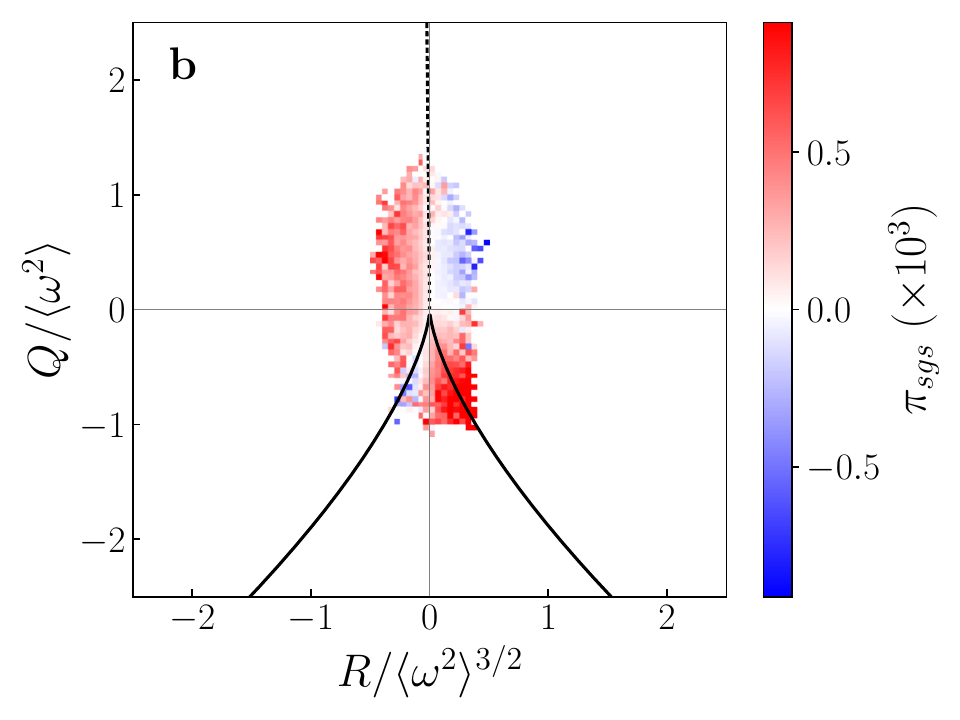}
    \caption{Joint PDFs of $Q$ and $R$ GTGIs (a) and averaged energy transfer rate in the $RQ$-plane (b) for $P\simeq 0$. GTGI values are normalized to powers of the enstrophy $\langle\omega^2\rangle$. Thick black line indicate the discriminant for $\bar{P}_0=-1.2\times10^{-3}$ and the dotted line indicate the $\bar{P}_0 Q-R=0$ curve.}
    \label{fig:pqr_pi_p_eq_0}
\end{figure*}
\begin{figure*}
    \centering
    \includegraphics[width=0.49\linewidth]{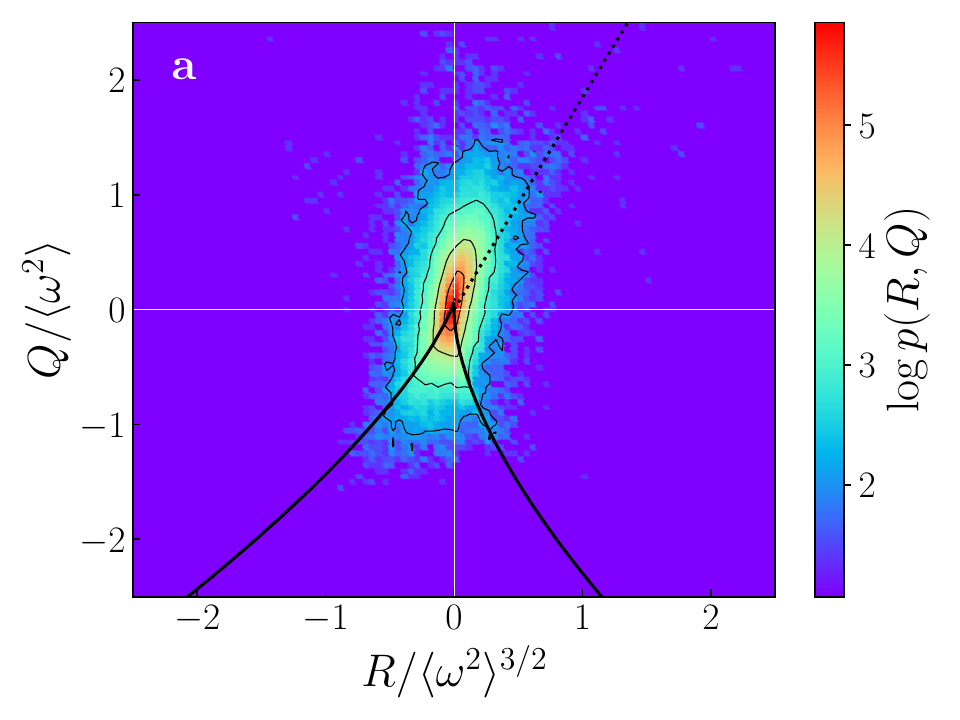}
    \includegraphics[width=0.49\linewidth]{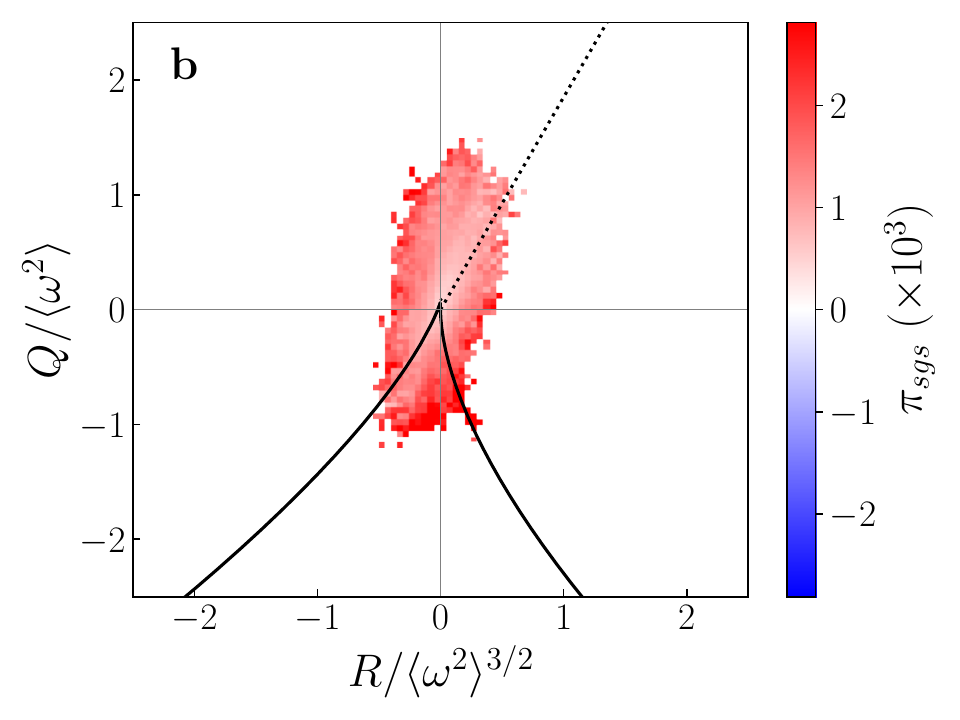}
    \includegraphics[width=0.49\linewidth]{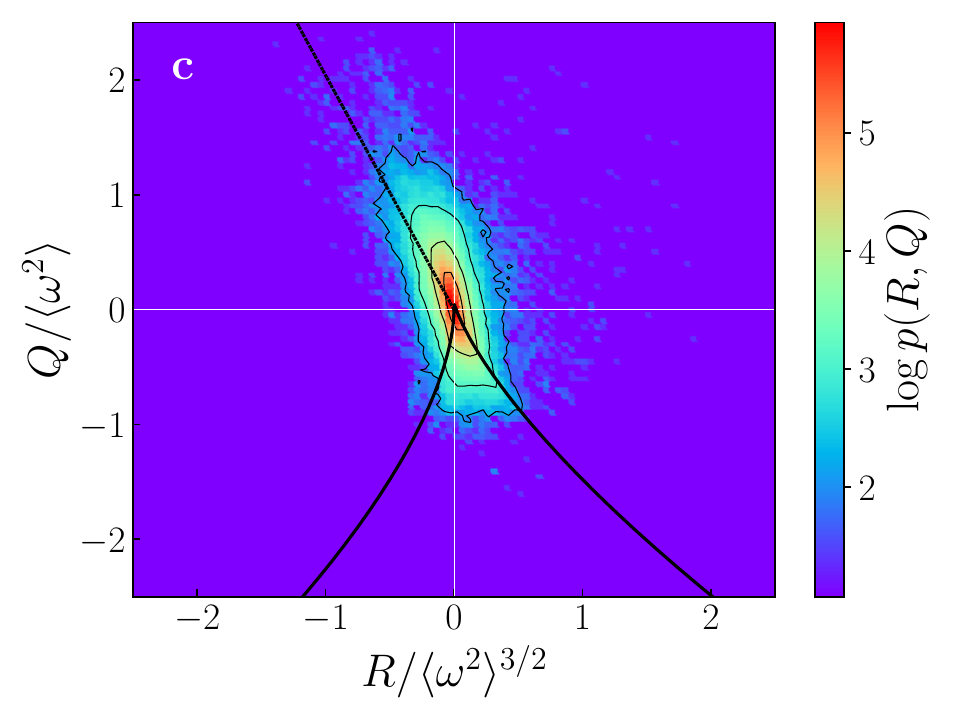}
    \includegraphics[width=0.49\linewidth]{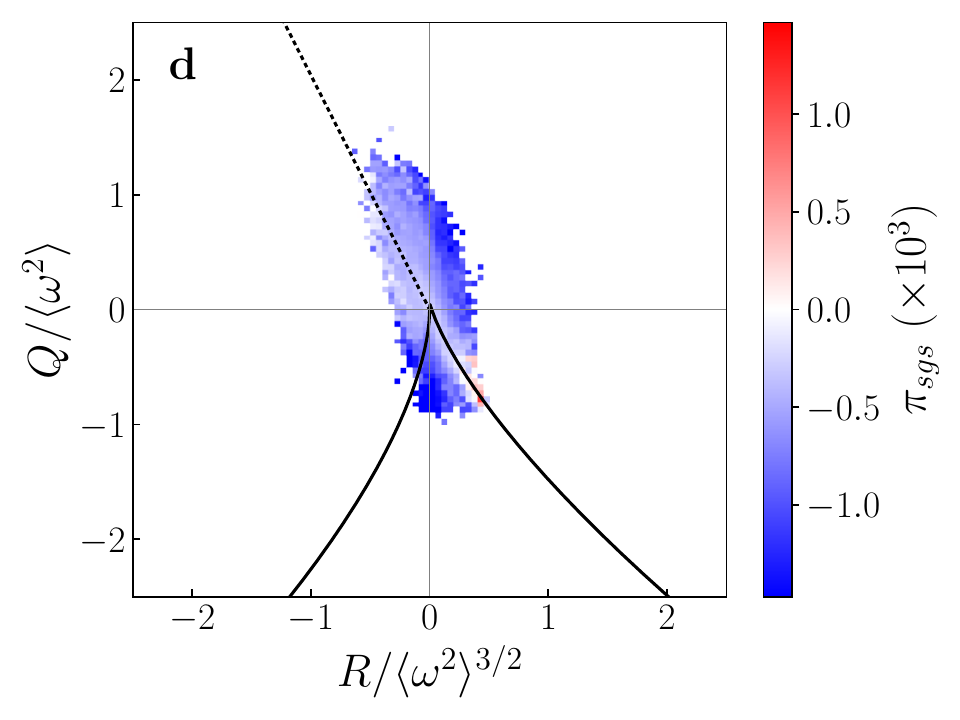}
    \caption{Joint PDFs of $Q$ and $R$ GTGIs (a) and averaged energy transfer rate in the $RQ$-plane (b) for $P\gg 0$. Thick black lines is the discriminant line obtained for $\bar{P}_\gg=0.07$ and the dotted line indicate the $\bar{P}_\gg Q-R=0$ curve. Joint PDFs of $Q$ and $R$ GTGIs (c) and averaged energy transfer rate in the $RQ$-plane (d) for $P\ll 0$. Thick black lines is the discriminant line obtained for $\bar{P}_\ll=-0.06$ and the dotted line indicate the $\bar{P}_\ll Q-R=0$ curve. GTGI values are normalized to powers of the enstrophy $\langle\omega^2\rangle$.}
    \label{fig:pqr_pi_p_neq_0}
\end{figure*}

We analyze a three-dimensional box of a Hall-MHD simulation with $b_{rms}=u_{rms}=0.25$, statisticially vanishing $u$-$b$ correlations, and resolution of $d_i/8$ at a fixed time. The numerical setup is described in Appendix \ref{sec:app}. To disentangle the different properties of compressible vs incompressible fluctuations we first present the statistics of the first velocity GTGI $P$, which is defined as $-\nabla\cdot\bm{u}$, see Equation \eqref{eq:p}. The probability distribution function (PDF) is reported in Figure \ref{fig:p_stat}. The obtained PDF is peaked around $P\sim0$ and is slightly skewed towards positive values of the invariant ($skew=0.23$). We divided the PDF in three different regions: one representative of nearly-incompressible fluctuations, and two comprising the most compressible fluctuations in the negative and positive tails of the PDF. Such thresholds have been set as $P\ll0$ below the 5\% percentile of the PDF, $P\gg0$ above the 95\% percentile and $P\sim0$ between 47.5\% and 52.5\% percentile, as indicated by the shaded areas in Figure \ref{fig:p_stat}, with the same sample size for the three cases. %The partition is unevenly chose, with the extreme tails ($P\ll0$ and $P\gg0$) accounting only for the 5\% of the total dataset, while the central region ($P\sim0$) comprises the 50\% of data. This choice is motivated by the need to isolate the most extreme compressibile fluctuations, and to retain a statistically robust subset of fluctuations with vanishing mean divergence to characterize the nearly-incompressible regions.

To evaluate the energy transfer rate we need to set a scale $\ell$ for CG the fields. Hereafter we consider $\ell=d_i$ as a characteristic scale, where the turbulent cascade is transferring energy from scales greater than $d_i$, and Hall term effects as well as energy dissipation are contained at scales smaller than $d_i$. In this fashion we can evaluate the direct transfer of energy coming from the turbulent cascade towards sub-ion scales, and also the local energy coming from subgrid scales associated with sub-ion dynamics and Hall effect. Starting from the incompressible case, we report the joint PDF of $Q$ and $R$ evaluated after the CG of the velocity field at $\ell=d_i$ in Figure \ref{fig:pqr_pi_p_eq_0}a. Although we filtered out the smallest-scale dissipative dynamics, the joint PDF still exhibits a notable asymmetry, with the tendency to develop a tail towards the $Q<0$ and $R>0$ part of the plane along the discriminant line $\Delta|_{P=\bar{P}_0}$, Equation \eqref{eq:d_p_neq_0}. Since $\bar{P}_0=-1.2\times10^{-3}$ is quite small, by calculating the separatrix between stable and unstable topologies $\bar{P}_0Q-R=0$ we obtain a curve very close to $R=0$ that corresponds to the incompressible case (Figure \ref{fig:pqr_pi_p_eq_0}). We then estimated the total energy transfer rate $\pi_{sgs}=\pi_{sgs}^\text{RM}+\pi_{sgs}^\text{BPY}+\pi_{sgs}^\text{MHD}+d_i\pi_{sgs}^\text{Hall}$ bin-averaged on the $RQ$-plane at $\ell=d_i$. Results are shown in Figure \ref{fig:pqr_pi_p_eq_0}b. Only bins with more than 10 counts are considered in the joint PDF. We observe a strong self-organization of the turbulent energy flux with respect to the local streamline topology. A direct energy cascade organizes in a topological selective fashion on compressing strain (sheet-like) structures, and stable stretching vortical structures. Conversely, negative transfer rate has been found to be mostly associated with unstable compressing vortical structures. These results closely resemble recent findings in numerical simulations of hydrodynamic turbulence reported by \cite{Yao2024}.

If we consider only compressible fluctuations, however, such a scenario changes. In Figure \ref{fig:pqr_pi_p_neq_0}a-b we report the joint PDF of $Q$ and $R$ and the transfer rate $\pi_{sgs}$ for the case $P\gg0$. The joint PDF in this case is distorted if compared with the $P\sim0$ case, following the discriminant line $\Delta|_{P=\bar{P}_\gg}$, where $\bar{P}_\gg=0.07$ has been chosen to be the median value of the right tail in the $P\gg0$ region, as shown in Figure \ref{fig:p_stat}. The energy transfer rate bin-averaged in the $RQ$-plane mostly develop positive values with a net weakening of the topological selection observed in the incompressible case (i.e., a change of the topology in the $RQ$-plane does not correspond to a reversal of the energy transfer). The minimum values of $\pi_{sgs}$ are organized along the $PQ-R=0$ line separating unstable compressing vortical structures (1) from volume contracting topologies (5). Analogous, though opposite, results are obtained in the $P\ll0$ case. In Figure \ref{fig:pqr_pi_p_neq_0}c-d we show the joint PDF of $Q$ and $R$ and the transfer rate $\pi_{sgs}$, and the PDF is distorted in the opposite sense with respect to the $P\gg0$ case. The shape of the PDF follows the discriminant line $\Delta|_{P=\bar{P}_\ll}$, Equation \eqref{eq:d_p_neq_0}, where $\bar{P}_\ll=-0.06$ is the median value of the left tail in the $P\ll0$ region showed in Figure \ref{fig:p_stat}. The opposite behavior is also observed in terms of total transfer rate, which tends to develop an inverse transfer almost everywhere on the plane, with a clear weakening along the $PQ-R=0$ separatrix. We also observe nearly-vanishing values of $\pi_{sgs}$ along the Vieillefosse tail, with even a few slightly positive bins.
In compressible regions, the dominance of the cascade sign by $P$ can be interpreted as a geometric effect: the local contraction (expansion) of the volume containing a certain amount of energy shifts this energy toward smaller (larger) scales. In other words, the volumetric compression itself acts as a cross-scale transfer mechanism.

\section{Discussions and conclusions}
In this letter we elucidated the connection between local streamline topology and subgrid-scale energy transfer rate in both incompressible and compressible plasma regions. In nearly incompressible regions, the SGS energy transfer self-organizes with the streamline topology in the $RQ$-plane: strain-compressing and vortical-stretching structures are associated with direct transfer rate towards small scales, whereas strain-stretching and vortical-compressing structures show a net back-transfer of energy from small scales. These results are shown for the particular choice of $\ell=d_i$, but similar properties can be also observed when moving towards inertial or sub-ion scales. As a matter of fact, the exact calculation of the transfer rate $\pi_{sgs}$ in the space-filter approach is not based on Yaglom's law or other models, and can be properly applied at any CG scale \citep{Aluie2017,Manzini2022}. Our results clearly highlight the primary role played by the velocity field in driving the LET rate, which nicely organizes in a topological selective fashion for nearly-incompressible fluctuations.

Concerning compressible subsets, the self-organization weakens and the sign of the transfer becomes primarily controlled by the local volumetric compression ($P>0$) or expansion ($P<0$), yielding predominantly forward transfer ($\pi_{sgs}>0$) in compressing regions and predominantly inverse transfer ($\pi_{sgs}<0$) on expanding sites. Although the sign of the first invariant $P$ determines the net volumetric compression or expansion, it does not uniquely determine the local streamline topology, which depends upon individual eigenvalues. Hence, compressing and expanding subsets are still characterized by a co-existence of different topological configurations, such types (1) to (4) of Figure \ref{fig:pqr}, although the direction of the energy transfer is primarily controlled by the sign of $P$.
Previous studies have shown that at kinetic scales bulk and magnetic energy conversion are driven by the pressure-strain interaction \citep{Yang2016,Yang2017,DelSarto2018,Matthaeus2020,Pezzi2021,Yang2022,Roy2025,Hellinger2025} in which  the compressive part typically plays a subdominant role. Here we provide a complementary information by showing that where compressive fluctuations become significant at MHD/ion scales, they act as a cross-scale transfer mechanism for turbulent energy. %A consequence of the relationship we have identified between $\pi_{sgs}$ and $P$ is that the turbulent energy transfer in average is directed towards small scales (i.e., $\pi_{sgs}>0$), that implies a  %From the definition of the first GTGI follows that $P=-\nabla\cdot\bm{u}$, hence we can state that all the region of spaces where velocity field is compressed ($\nabla\cdot\bm{u}\ll0$) experience a transfer of energy towards small scales. On the contrary, a back-transfer of energy is observed in all the locally expanding region of the simulation, i.e., where $\nabla\cdot\bm{u}\gg0$.

From a broader perspective, our findings indicate that the velocity GTGIs contain predictive information on the sign of the energy flux for nearly-incompressible fluctuations, that can be linked to the co-existence of direct and inverse energy transfer routinely observed in nearly-incompressible Alfvénic streams \citep{Smith2009,Coburn2014}. %An important implication for space and astrophysical plasmas is that the direction of the cross-scale energy transfer is primarily controlled by the local volumetric compression/expansion, rather than by the detailed streamline topology, and the spatial distribution of the transfer rate may be largely organized by high compressional structures, such as shocks, strongly influencing the energy conversion \citep{Yang2016}. Our 
The strong correlation observed between streamline topology and energy transfer rate closely resembles previous observations in hydrodynamic turbulence in the incompressible case \citep{Bos2002,Yao2024}, extending them to the case of Hall-MHD plasma turbulence in the more general and still unexplored compressible case. In compressible regions, indeed, we observe a weakening of the topological selection of the energy transfer, since the overall sign of $\pi_{sgs}$ is determined by the sign of $P$ (or equivalently $-\nabla\cdot\bm{u}$). %This result represents a new piece of information complementing previous works that demonstrated how compressive effects control the energy conversion and dissipation in plasma turbulence 

%We highlighted the interplay between the streamline configurations and energy transfer and dissipation in plasma turbulence by calculating 
%\begin{itemize}
%    \item the SGS energy transfer terms on the $R_A$-$Q$ invariant plane;
%    \item the viscous and ohmic dissipation terms at different scales.
%\end{itemize}
%We found that the SGS energy transfer terms exhibit a distinct pattern as a function of the streamline topology. A direct cascade is observed by investigating the Reynolds-Maxwell and the MHD terms, and it is clustered around two types of structures, i.e., stable vortices and strain-dominated structures, whereas unstable vortices are associated with an inverse energy transfer. Conversely, the Hall SGS term displays an opposite behavior with a backscattering of energy from small scales associated with strain and stable vortical structures and a direct energy transfer in correspondence with unstable vortices. This represents an interesting observation, possibly reflecting the non-dissipative nature of the Hall term, which acts as a balance and/or back-reaction with respect to the MHD term at ion scales.

%Going towards the smallest scales of the energy dissipation, we calculated the viscous and ohmic terms and we investigated how they clusterize in the $R$-$Q$ plane at the grid-resolution scale and at larger scales by CG the fields. Our results suggest a self-organization of the local configuration of the field lines in the cascading process, since a net correlation between the pattern observed at the dissipative scales persists up to inertial scales.

The study of the cascade rate associated with field line topology from an experimental point of view needs a multipoint constellation, and can be assessed in principle through missions like Cluster or MMS. Very recently, this kind of effort has been made by \cite{Hnat2025} on Cluster data, and authors conclude that magnetic field topology plays a role in the energy transfer rate by showing that a positive cascade rate is found to be associated with a predominance of strain-dominated magnetic field structures. This result supports the importance of reconnecting current sheets in driving the energy cascade towards small scales, also supported by several numerical studies \citep{Franci2017,Manzini2023,Foldes2024}. %However, our results reveal an even more significant correlation between local energy cascade and streamline topology, and %A trace of this transfer rate can be observed also by inspecting the cross-coupling between ion-scale structures and the viscous dissipation as computed at the grid resolution. Indeed, whereas in principle we can expect to find the viscous dissipation completely clustered below the discriminant line, being defined in terms of self contraction the strain tensor, we observe a distinct dissipative pattern in correspondence with stable vortical structures. This correlation deserves theoretical investigation which is here left as a future work. The weak correlation between cascade rate and magnetic field local topology can be associated with a different, possibly smaller, reconnection rate with respect to Cluster data analyzed by \cite{Hnat2025}. 
A dedicated investigation of the specific role of magnetic reconnection in the self-organization of magnetic field line structures from numerical simulations is here left as a future investigation.

%Previous studies have shown that at kinetic scales bulk and magnetic energy conversion are driven by the pressure-strain interaction \citep{Yang2016,Yang2017,DelSarto2018,Yang2022,Roy2025,Hellinger2025} in which  the compressive part typically plays a subdominant role. Here we provide a complementary  information by showing that where compressive fluctuations become significant they act as the main driver of the turbulent energy transfer. In this sense, the local volumetric compression acts as a cross-scale transfer mechanism. %The relevance of our results lies in the fact that positive and negative energy cascade rates can be linked to different classes of local plasma structures. In other words, the GTGI analysis allows us to further discriminate which specific topological configuration is associated with direct or inverse local cascade rate. While previous studies have identified coherent structures such as current sheets, Alfvén vortices, and compressive magnetic structures as key elements of intermittency, energy transfer, and dissipation \citep[e.g.,][]{Marsch1997,Alexandrova2006,Servidio2008,Greco2009,Matthaeus2015,Perrone2016,Perrone2017}, t

%In this letter, we highlight the importance of including the velocity-field GTGIs as an additional and complementary diagnostic to LET, extending the toolbox routinely employed in the analysis of turbulence in space plasmas, with relevance for future spacecraft constellations. In fact, a
All the dynamics of the energy cascade illustrated in this letter would greatly benefit from the simultaneous exploration of different spatial scales. The only way to test our results across different scales is to develop future multiscale constellation such as Helioswarm \citep{Klein2023} and the proposed Plasma Observatory mission \citep{Rae2022,Retino2022}, where different scales in heliospheric and magnetospheric plasmas can be simultaneously accessed in situ. In the context of a growing fleet of future multipoint and multiscale constellations, we emphasize the fundamental importance of including the study of the streamline topology by means of GTGI as a standard diagnostic for plasma turbulent studies. Indeed, whereas LET can give important insight about the local direction only of the energy cascade, the investigation of GTGIs adds the fundamental piece of information about the type of structures that are locally transferring the energy, thus enabling a more comprehensive view of turbulent cascade in space plasmas.

\begin{acknowledgments}
This work has been partially supported by Agenzia Spaziale Italiana (ASI) under the agreement ASI-INAF ``Attività di Fase A per la missione Plasma Observatory'' F83C24000690001. A.V. and G.C acknowledge partial financial support from the European Union – Next Generation EU – National Recovery and Resilience Plan (NRRP) – M4C2 Investment 1.1- PRIN 2022 (D.D. 104 del 2/2/2022) – Project “Modeling Interplanetary Coronal Mass Ejections”, MUR code 31. 2022M5TKR2, CUP B53D23004860006 and C53D23001190006.
We acknowledge partial funding by “Fondazione Cassa di Risparmio di Firenze” under the
project HIPERCRHEL. E.P. acknowledges CINECA and
INAF for awarding access to HPC resources under the coordination of the "Accordo Quadro MoU per lo svolgimento di attività congiunta di ricerca Nuove frontiere in Astrofisica: HPC e Data Exploration di nuova generazione" (project INA23\_C9A10).
\end{acknowledgments}

%\begin{contribution}
%All authors contributed equally to the Terra Mater collaboration.
%\end{contribution}

\bibliography{gtgi_apj}{}
\bibliographystyle{aasjournalv7}

\appendix
\section{Coarse-grained compressible Hall-MHD equations}\label{app:hmhd_cg}
In this work we use the SGS energy transfer terms arising from the CG of the compressible Hall-MHD equations. In this section we summarize the derivation of all the energy transfer terms denoted as $\pi_{sgs}$. The set of compressible Hall-MHD equations comprises the continuity equation
\begin{equation}
    \partial_t\rho+\partial_j(\rho u_j)=0,
\end{equation}
the momentum equation
\begin{equation}
    \partial_t(\rho u_j)+\partial_k\biggl[\rho u_ju_k + \biggl(p+\frac{b^2}{2} \biggr)\delta_{jk}-b_jb_k \biggr]=\mathcal{D}_j^{(u)},
\end{equation}
and the induction equation
\begin{equation}
    \partial_tb_j=\epsilon_{jkl}\partial_k\biggl[\epsilon_{lmn}u_mb_n-d_i\epsilon_{lmn}\frac{J_mb_n}{\rho} \biggr]+\mathcal{D}_j^{(b)}.
\end{equation}
In these equations $d_i$ is the ion inertial length, and $\mathcal{D}_j^{(u)}$, $\mathcal{D}_j^{(b)}$ indicate the dissipative terms that are not explicitly written since they do not contribute to the SGS energy transfer. By filtering the equations with a Gaussian kernel, we indicate with $\bar{f}$ the filtered version of the field $f$ and with $\tilde{u}_i=\overline{\rho u_i}/\bar{\rho}$ the Favre filtered version of the velocity field. We then define the stress tensor $\bar{\tau}(f,g)=\overline{fg}-\bar{f}\bar{g}$ along with its Favre filtered version for the velocity field $\tilde{\tau}(u_i,u_j)=\widetilde{u_iu_j}-\tilde{u}_i\tilde{u}_j$. The coarse-grained versione of the compressible Hall-MHD equations are
\begin{equation}
    \partial_t\bar{\rho}+\partial_j(\bar{\rho}\, \tilde{u}_j)=0
\end{equation}
for the continuity equation, then
\begin{equation}
    \partial_t(\bar{\rho}\,\tilde{u}_j)+\partial_k\biggl[\bar{\rho}\,\tilde{u}_j\tilde{u}_k+\biggl(\bar{p}+\frac{\bar{b}^2}{2}\biggr)\delta_{jk}-\bar{b}_j\bar{b}_k \biggr]=-\partial_k[\bar{\rho}\,\tilde{\tau}(u_j,u_k)-\bar{\tau}(b_j,b_k)]+\overline{\mathcal{D}}_j^{(u)}
    \label{eq:u_cg}
\end{equation}
for the momentum equation, and
\begin{equation}
    \partial_t\bar{b}_j=\epsilon_{jkl}\partial_k\biggl[\epsilon_{lmn}\biggl(\tilde{u}_m\bar{b}_n-d_i\frac{\bar{J}_m\bar{b}_n}{\bar{\rho}}\biggr)+\mathcal{E}_l^\text{MHD}-d_i\mathcal{E}_l^\text{Hall} \biggr]+\overline{\mathcal{D}}_{j}^{(b)}
    \label{eq:b_cg}
\end{equation}
for the induction equation. The terms $\mathcal{E}_l^\text{MHD}=\overline{\epsilon_{lmn}u_mb_n}-\epsilon_{lmn}\tilde{u}_m\bar{b}_n$ and $\mathcal{E}_l^\text{Hall}=\overline{\epsilon_{lmn}J_mb_n/\rho}-\epsilon_{lmn}\bar{J}_m\bar{b}_n/\bar{\rho}$ indicate the MHD and Hall electromotive forces, respectively. By multiplying  Equation \eqref{eq:u_cg} for $\tilde{u}_j$ and Equation \eqref{eq:b_cg} for $\bar{b}_j$, we can calculate the rate of change of kinetic and magnetic energy, and by introducing the term $\mathcal{F}_j$ containing all the contributions that can be written as spatial fluxes (i.e., written in divergence form), we obtain 
%\begin{equation}
    %\partial_t\bar{E}_u+...
%\end{equation}
%and
%\begin{equation}
    %\partial_t\bar{E}_b+...
%\end{equation}
%Finally, 
the total energy balance as
\begin{equation}
    \partial_t(\bar{E}_u+\bar{E}_b)+\partial_k\mathcal{F}_k=-\bar{p}\,\partial_j\bar{u}_j-\pi_{sgs}+\tilde{u}_j\overline{\mathcal{D}}_j^{(u)}+\bar{b}_j\overline{\mathcal{D}}_j^{(b)},
\end{equation}
where the spatial flux term is
\begin{multline}
    \mathcal{F}_k=\bar{E}_{u}\tilde{u}_k+\bar{p}\bar{u}_j\delta_{jk}+\frac{\bar{b}^2}{2}\tilde{u}_j-\tilde{u}_j\bar{b}_j\bar{b}_k-d_i\epsilon_{jkl}\frac{\epsilon_{lmn}\bar{J}_m \bar{b}_n}{\bar{\rho}}\bar{b}_l\\+\bar{\rho}\tilde{u}_j\tilde{\tau}(u_j,u_k)-\tilde{u}_j\bar{\tau}(b_j,b_k)-\bar{J}_k(\mathcal{E}^\text{MHD}_k+d_i\mathcal{E}^\text{Hall}_k)
\end{multline}
and the subgrid energy transfer term is
\begin{multline}
    \pi_{sgs}=-[\bar{\rho}\,\tilde{\tau}(u_j,u_k)-\bar{\tau}(b_j,b_k)]\,\partial_k\tilde{u}_j-\frac{1}{\bar{\rho}}\,\partial_j \bar{p}\,\bar{\tau}(\rho,u_j)-\mathcal{E}^\text{MHD}_j\,\bar{J}_j-d_i\mathcal{E}^\text{Hall}_j\,\bar{J}_j\\=\pi_{sgs}^\text{RM}+\pi_{sgs}^\text{BPY}+\pi_{sgs}^\text{MHD}+d_i\pi_{sgs}^\text{Hall}.
\end{multline}
Superscript indices refer to Reynolds-Maxell (RM) stress tensor, baropycnal (BPY) work, MHD, and Hall SGS energy transfer rate.

\section{Numerical Setup}\label{sec:app}
\begin{figure}
    \centering
    \includegraphics[width=0.5\linewidth]{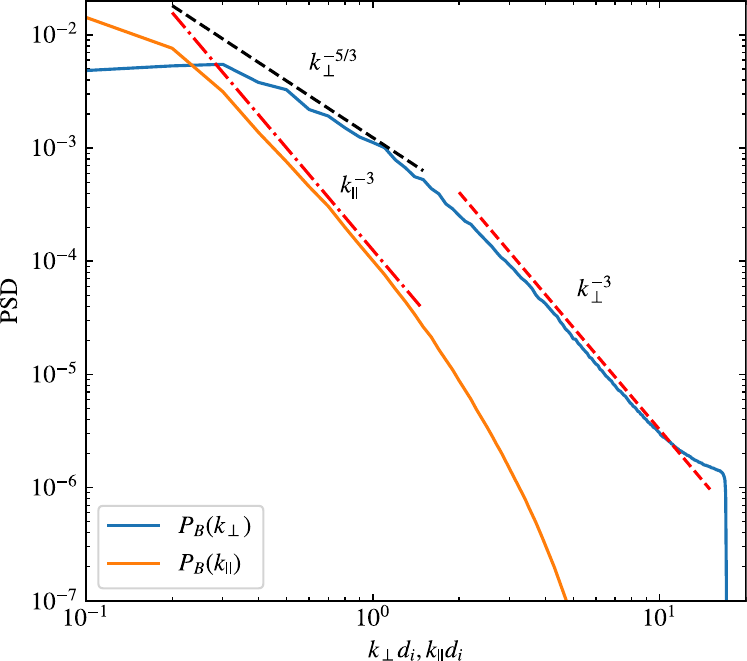}
    \caption{Reduced 1D perpendicular and parallel spectra of magnetic field fluctuations w.r.t the mean magnetic field. Dashed and dot-dashed lines represent typical slopes expected in inertial and Hall ranges that serve as reference.}
    \label{fig:hmhd_spectrum}
\end{figure}

In this work, we employ a pseudospectral code that solves the compressible viscous-resistive Hall-MHD equations \citep{Montagud2021}. 
%\begin{align}
%\partial_t \rho &= -\nabla(\rho\mathbf{u})\\
%\rho D_t\mathbf{u} &= -\nabla P + (\nabla\times\mathbf{b})\times\mathbf{b}
%+ \mu\!\left[\nabla^2\mathbf{u}+\frac{1}{3}\nabla(\nabla\!\cdot\!\mathbf{u})\right]\\
%D_t P &= -\gamma P\,\nabla\!\cdot\!\mathbf{u}
%+(\gamma-1)\rho\eta|\nabla\times\mathbf{b}|^2\\
%&\quad+(\gamma-1)\mu\!\left[|\nabla\times\mathbf{b}|^2
%+ \frac{4}{3}(\nabla\!\cdot\!\mathbf{b})^2\right]\\
%\partial_t\mathbf{b} &= \nabla\times(\mathbf{u}\times\mathbf{b})
%+ \eta\nabla^2\mathbf{b}
%- d_i\frac{\nabla\times[(\nabla\times\mathbf{b})\times\mathbf{b}]}{L},
%\label{eq:numset}
%\end{align}
%where $D_t = \partial_t + \bm{u}\cdot\nabla$ is the lagrangian derivative and $\gamma = 5/3$ is the adiabatic index. 
In the code, all physical quantities are normalized with respect to a characteristic length $L = d_i$ set to be equal to the ion inertial length, a plasma density $\rho_0$, a magnetic field amplitude $B_0$, and a pressure $P_0 = \rho_0 c^2_A$ (being $c_A=B_0 / \sqrt{4\pi\rho_0}$ the Alfvén speed). Dynamic viscosity $\mu$ and magnetic resistivity $\eta$ are in units of $d_i c_A \rho_0$ and $d_i c_A$ respectively. The simulation box consists of a cube with periodic boundaries whose edge is $20\pi d_i$ long. A grid of $512^3$ points is employed, corresponding to a spatial resolution of 0.1227 $d_i$ in all three directions. There is a mean background magnetic field along $\hat{\bm{x}}$, i.e., $\bm{B}_0 = B_0\, \hat{\bm{x}}$. 
The simulation is initialized with a spectrum of fluctuations with constant amplitudes, random phases, and zero-mean cross-helicity \citep{Franci2018}. The amplitudes of magnetic and velocity fluctuations are set such that the root mean square of the fluctuations is  $B_{rms} = u_{rms} = 0.25$. 
Finally, we set the plasma $\beta = 2P_0 /B^2_0 = 2$, $\nu = \eta = 10^{-3}$, corresponding to a Reynolds number of 4000. 
The simulation evolves as a classic Alfvenic decaying turbulence simulation. The dataset employed in this work corresponds to a snapshot of the simulation taken at the time whn turbulence has fully developed. The reduced 1D power spectra of magnetic field fluctuations are displayed in Figure \ref{fig:hmhd_spectrum}, and give similar spectral properties to those reported by \citet{Franci2018} for 3D hybrid numerical simulations of turbulence. In particular, the parallel 1D reduced spectrum exhibit a slope $\sim k_\parallel^{-3}$, while the reduced 1D perpendicular spectrum shows an inertial range, $\sim k_\perp^{-5/3}$ below $k_\perp d_i=1$, and then develops a $\sim k_\perp^{-3}$ trend at sub-ion scales.

\end{document}